\documentclass{IEEEcsmag}

\usepackage[colorlinks,urlcolor=blue,linkcolor=blue,citecolor=blue]{hyperref}

\usepackage{upmath}

\jvol{XX}
\jnum{XX}
\paper{8}
\jmonth{May/June}
\jname{IT Professional}
\pubyear{2019}

\begin{document}

\sptitle{Department: Head}
\editor{Editor: Name, xxxx@email}

\title{Security of Deep Learning Methodologies: Challenges and Opportunities
%An Overview
}

\author{Shahbaz Rezaei}
\affil{University of California, Davis}

\author{Xin Liu}
\affil{University of California, Davis}

\markboth{Department Head}{Paper title}

\begin{abstract}
Despite the plethora of studies about security vulnerabilities and defenses of deep learning models, security aspects of deep learning methodologies, such as transfer learning, have been rarely studied. In this article, we highlight the security challenges and research opportunities of these methodologies, focusing on vulnerabilities and attacks unique to them.
\end{abstract}

\maketitle

\textbf{Keywords:} Machine Learning Security; Privacy; Deep Learning; Machine Learning Methodologies.

\chapterinitial{W} 
ith the widespread adaptation of deep neural networks (DNN), their security challenges have received significant attention from both academia and industry, especially for mission critical applications, such as road sign detection for autonomous vehicles, face recognition in authentication systems, and fraud detection in financial systems.

%need to be studied and addressed to have a safe and ongoing future growth. In particular, certain tasks are more mission critical and need stronger security considerations, e.g., road sign detection for autonomous vehicles, face recognition in authentication systems, and fraud detection in financial systems.

%In the past few years, many studies have shown that deep learning models suffer from several types of attacks, such as adversarial attacks, data poisoning, and exploratory attacks. 
There are three major types of attacks on deep learning models, namely adversarial attacks, data poisoning, and exploratory attacks.
Particularly, adversarial attacks, which aim to carefully craft inputs that cause the model to misclassify, has been extensively studied and many defence mechanisms have been proposed to alleviate them. These attacks are of paramount importance because they are effective, moderately simple to launch, and often transferable from one model to another. In literature, there are several survey and review papers on deep learning security and defence mechanisms. In this article, we focus on security of a much less explored area of machine learning - machine learning {\it methodologies.}

Machine learning methodologies have been widely used to mitigate the restrictions and assumptions of a typical machine learning process. A typical DNN training process assumes large labeled dataset(s), access to high computational resources, non-private and centralized data, standard training and hyper-parameter tuning, and fixed task distribution over time. However, these assumptions are often difficult to realize in practice. As a result, different machine learning methodologies have been developed and adopted, such as transfer learning, federated learning, model compression, multi-task learning, meta-learning, and lifelong learning. Notwithstanding the proliferation of these machine learning methodologies, their security aspects have not been comprehensively analyzed, if ever studied.

In this article, we focus on potential attacks, security vulnerabilities, and future directions specific to each learning methodology. Note that there are many more machine learning methodologies in literature, including few-shot learning, on-device learning, zero-shot learning, to name but a few. However, due to the lack of space and the fact that these methodologies mostly overlap with the ones that we review, we limit our discussion to the aforementioned methodologies. We assume that readers have rudimentary background on deep neural networks and how they work.

\section{Background}

\subsection{Attack Taxonomy}

In machine learning security, an attack has a threat model that defines the goal, capabilities (or knowledge), and target model. The attacker's goal can be categorized in terms of security violation: 1) violation of \textit{availability} that aims to reduce the confidence of a model for normal inputs, 2) violation of \textit{integrity} that aims misclassification on certain inputs without affecting normal inputs, and 3) violation of \textit{privacy} that aims to obtain confidential information about the model, training or inference-time data and users, or even hyper-parameters used during training (hyper-parameter stealing attack).

The life-cycle of a typical machine learning model with offline training data consists of training and inference phases, which indicate attacker's capabilities and knowledge. Training phase capabilities are \textit{data injection}, where the attacker injects new data points to the training dataset, \textit{data poisoning}, where the attacker modifies the existing data points in the training dataset, and \textit{logic corruption}, where the attacker interferes with the learning algorithm. 

In the inference phase, the model is assumed to be fixed and the attacker cannot change the model. However, the attacker can still craft data inputs that fool the model to provide incorrect outputs. Hence, the attacker's capability is defined based on how much information she has about the model, ranging from \textit{white-box}, where everything is possibly known including the entire model and training data, to \textit{black-box} attacks, where minimum knowledge about the model, training data and algorithm is known.
Any attack model that lays between while-box and black-box attack in terms of available information about the model is called \textit{gray-box} attack.

\begin{figure*}
\centerline{\includegraphics[width=1\linewidth]{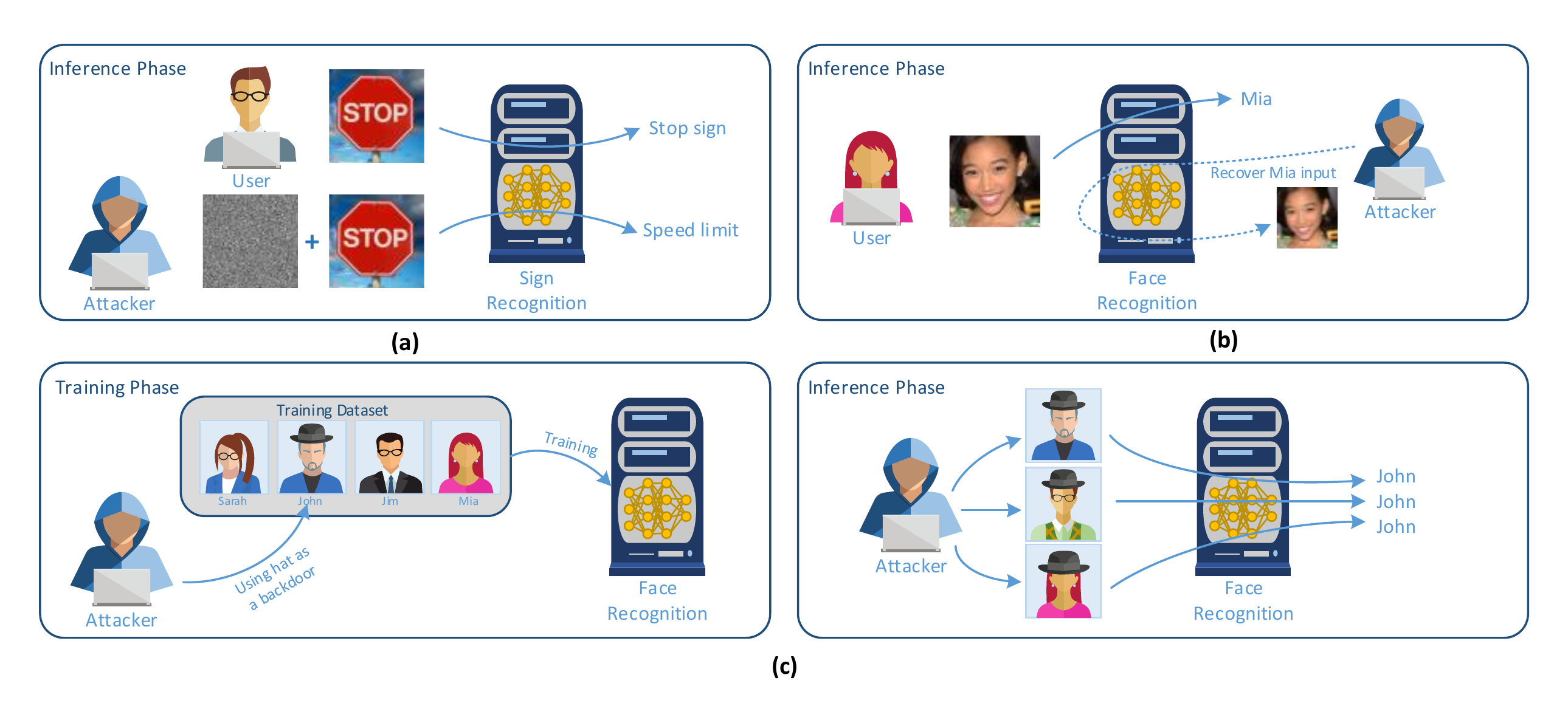}}
\caption{Typical attacks on machine learning: (a) Adversarial attack, (b) model inversion attack, and (c) backdoor attack.}
\label{fig-attacks}
\end{figure*}

\subsection{Attack Types}
In machine learning security, attacks are often categorized into three attack types based on the threat model:

\textbf{Evasion attack (adversarial attack):} The goal of an evasion attack is to manipulate the input data such that the model misclassifies. Although one can technically manipulate training data using evasion attack methods during training phase (often for adversarial retraining as a defense mechanism), evasion attack is an inference phase attack that violates the integrity. Figure \ref{fig-attacks}(a) illustrates the adversarial attack where the attacker add a small perturbation, imperceptible to human eye, to the stop sign image to cause the model to misclassify.

\textbf{Data poisoning:} This is a training phase attack where the attacker inject or manipulate training data to either create a backdoor to use at inference time (without compromising the model performance on normal input data) or to corrupt the training process. Hence, it can violate availability or integrity depending on the goal. A typical example is to create a backdoor for face recognition task where the attacker injects a set of training samples with a specific object in a target person's training data. The aim is to force the model to associate the specific object with the target class. Then, any face image with the object is classified as the target class even if it belongs to another person. For instance, in Figure \ref{fig-attacks}(c), the attacker inject faces of John with a special hat during training. Then, at the inference phase, any face that has the hat is classified as John by the model.

\textbf{Exploratory attacks:} The aim of the attack is to violate the privacy at inference phase. It covers several types of attacks, including \textit{model extraction}, to extract model parameters, \textit{membership inference attack}, to examine whether a data point is used during the training phase, \textit{model inversion}, to infer something about input by observing the model output. In Figure \ref{fig-attacks}(b), the attacker aims to recover the input image of Mia by observing the output and the model. Although exploratory attacks have been widely studied for classical machine learning algorithms, there are only a few work on deep learning models. For example, it has been recently shown that sensitive and out-of-distribution sentences, such as "My social security is ----", can be leaked from commercial text-completion neural networks \cite{carlini2019secret}.

%\begin{figure*}
%\centerline{\includegraphics[width=0.8\linewidth]{methodologies-diagram}}
%\caption{Machine learning methodologies studied in this article.}
%\label{fig-methods}
%\end{figure*}

\subsection{Machine Learning Methodologies}
Several machine learning methodologies, such as tranfer learning or multi-task learning, have been used to train better models, reduce training data, reduce model complexity, distribute resources, or use other training data or models. These methodologies may overlap or have similarities. However, each methodology is based on different set of assumptions that may introduce different vulnerabilities. Here, we briefly explain the assumptions and goals of each machine learning methodology:

\textbf{Transfer learning:} Transfer learning refers to techniques that use the knowledge learned for one task, called the source task, to improve the performance of another task, called the target task. The model trained for the source task is called the teacher model and the model trained for the target task is called the student model. The most common way of transfer learning for deep learning models is to transfer all/some weights of a teacher model to a student model and then train the student model for the target task. The intuition is that the teacher model's weights are much closer to the local optimum for the target task than any random initialization of the student model. Hence, the training procedure of the model for the target task starts at the point where it is considerably close to the performance target of the training procedure.

\textbf{Multi-task learning:} The goal of multi-task learning is to learn several related tasks simultaneously. The most prevalent multi-task learning approach for deep models is parameter sharing, where the model consists of shared layers and task-specific layers. An example of multi-task learning is to predict the class and the coordinates of an object in an image.

\textbf{Federated learning:} Federated learning (FL) is a distributed learning approach that aims to train a model on a distributed private data. The assumption is that data is distributed among nodes capable of training a model, such as smartphone devices, and there is a centralized server that coordinates the training. The main approach consists of several training rounds where the server sends the model to a set of nodes. Then, these nodes train the model with all or a portion of their local dataset and send the updates back to the server. Then, at the end of each round, the server receives all the updates from the nodes and aggregates them to built a new model. The advantage of federated learning is that contributing nodes do not need to share or reveal their training data.

\textbf{Model compression:} The aim of model compression is to make large deep models suitable for devices with limited resources (e.g. CPU, memory, energy, bandwidth). The two common compression approaches are \textit{pruning} that reduces the number of parameters of models, and \textit{quantisation} that reduces the number of bits required to store each parameter. Although compression methods try to keep the model accuracy intact on the training data, the compressed model may act completely different from the original model on unseen or adversarial inputs.

\textbf{Meta-learning:}
Meta-learning is the process of better learning a task from past experiences and other tasks with the purpose of learning much faster, inspired by how humans learn. Meta-learning has been proliferated in the past two decades. Different categories and approaches of meta-learning is out of the scope of this paper. Avid readers can find several recent surveys on this topic. Note that meta-learning may overlap much with other methodologies, including transfer learning, few-shot learning, and multi-task learning. However, in this article, we consider broader and general categories of meta-learning.

\textbf{Lifelong machine learning:}
The ability to continually learn new tasks by building upon previously learned knowledge and tasks over time, which is an indispensable part of humans and animals life, is called lifelong learning. In lifelong learning, the system should retain its ability to perform older tasks and learn new tasks. lifelong learning has been studied for decades and the most difficult challenge is the catastrophic forgetting, which refers to the inability of models to perform old tasks as accurate as new tasks over time.

\section{Attacks on Deep Learning Methodologies}

\subsection{Transfer Learning}
\begin{figure}
\centerline{\includegraphics[width=1\linewidth]{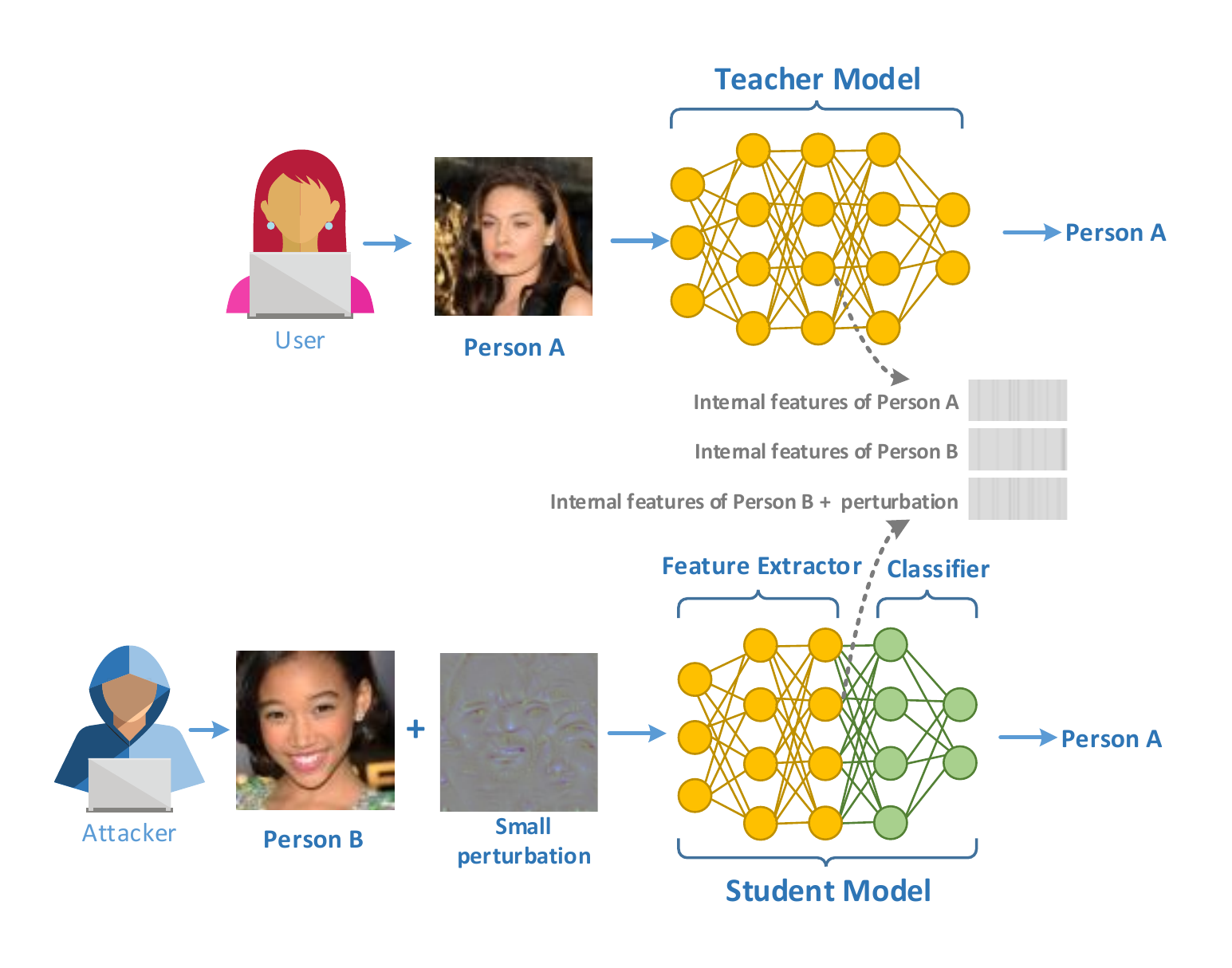}}
\caption{An adversarial attack on TL. The attacker that has the teacher model can use the feature extractor of the model to craft a small perturbation that changes the internal feature of the person B to be similar to that of person A. }
\label{fig-TL}
\end{figure}

The most common form of transfer learning (TL) is to transfer the first \textit{n} layers of the teacher model to the student model, add a few layers at the end of the model, and retrain the student model with a new dataset. As shown in Figure \ref{fig-TL}, the part transferred from the teacher model is called feature extractor and the new layers are called classifier. The feature extractor provides the high level representation of the input, such as the existence of certain objects in object detection, some frequencies and patterns in voice detection, etc.

The main vulnerability of TL methods stems from the fact the teacher models are often publicly available and known. Hence, even in the black-box scenario, an attacker can easily obtain the feature extractor part of the student model. It is shown that even if the entire student model is retrained with a new dataset, the feature extractor part is still close to the teacher's feature extractor \cite{wang2018great}. Furthermore, an attacker can manipulate the original teacher model and create a backdoor on it and upload it as a valid version of the teacher model. 

In literature, three studies developed strong attacks on TL. All the attacks assume white-box access to the teacher model, but black-box access to the student model. In \cite{wang2018great}, authors discuss that if two input images share similar internal representation, the classifier likely classifies them as the same class. Hence, they use the feature extractor from the teacher model to craft a small perturbation for the source input to be classified as another class. As shown in Figure \ref{fig-TL}, one can add a small perturbation to person B's image to fool the student model to classify it as person A. The optimization formulations used to craft adversarial inputs in all studies for machine learning strategies are similar to the formulation in attacks on deep models. Hence, we do not cover the optimization formulation in this article. 

Similar approach has also been used in \cite{ji2018model} in a form of data poisoning attack. They craft a set of adversarial images with similar internal features to the internal features of an attack target class. Then, they retrain the teacher model with the crafted images to create a backdoor. Hence, any student model that uses the poisoned model as a teacher is likely prone to the backdoor. It is challenging to detect models with backdoors since their behavior to the normal input is similar to the original model. Detecting backdoors of machine learning models is still an open problem.

In \cite{rezaei2019target}, the authors introduce a target-agnostic attack that does not need any sample input of the target class to trigger it. They show that even if one does not know the typical internal features of the target class for natural data, one can still trigger the Softmax layer with high probability. They argue that since the Softmax layer performs linear operation on the feature extractor, numerous internal features can produce the same output on the Softmax layer as the internal features of the target class have. Hence, they introduce a mechanism that attempts to craft images that has a very large value on one of the internal features. Such crafted images likely trigger the class that assigns a higher weight to that feature. This attack is even useful on systems that no input sample is available to the attacker, such as identification/authentication systems. However, they show that this target-agnostic attack can be defended by using a more sophisticated classifier that takes the distributional pattern of internal features into account, not just the linear combination.

Exploratory attacks are generally harder for deep learning models due to their complexity. There is no study on exploratory attack specifically for TL so far. However, due to the fact that the feature extractor part of the model is known to the attacker, exploratory attacks may be even easier. For instance, the model extraction attack, that aims to extract the entire model parameters, needs to only find the parameters of the classifier part. So, the search space is significantly smaller than the entire model parameters.

The main assumption that enables such attacks is that the teacher model is known and the feature extractor of the student model is not significantly different than the teacher model. Hence, the most effective mechanism that can potentially prevent all these attacks is to make a considerable change to the feature extractor so that it becomes different from the publicly available one. Unfortunately, the typical retraining process of a student model does not change the feature extractor significantly. Hence, in \cite{wang2018great}, they suggest a retraining optimization formulation with a constraint that forces the distance between the student feature extractor and the teacher's one to be higher than a threshold. Such an approach can reduce the chance of successful attack on TL. However, it increases the retraining time and requires the input samples and labels of the original teacher model.

\begin{figure*}
\centerline{\includegraphics[width=0.7\linewidth]{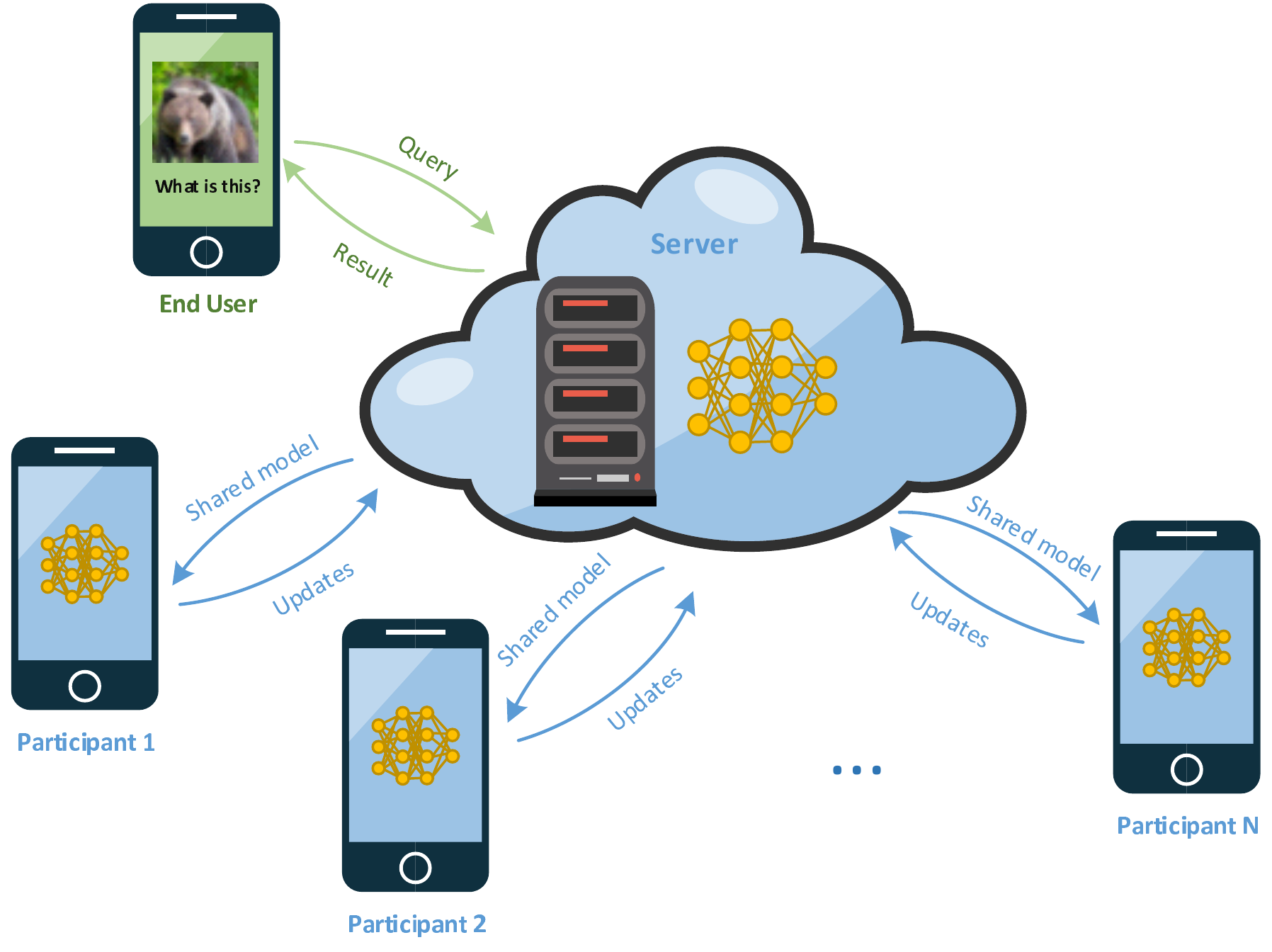}}
\caption{Federated learning components and training procedure.}
\label{fig-FL}
\end{figure*}

\subsection{Federated Learning}
Federated learning (FL) allows several participants to train a joint model using their local private data. As it is shown in Figure \ref{fig-FL}, FL consists of several rounds in which the shared model parameters are sent to a set of participants. Then, each participant updates the model parameters by training the model on a subset of local private data and send the parameter updates back to the server. Finally, the server aggregates all the updates and updates the shared model. The number of papers about security vulnerabilities and defences of FL methodology is disproportionally large compared to other methodologies. In this section, we briefly introduce the security vulnerabilities and defense types, and encourage readers to follow the references and related papers for more details.

As shown in Figure \ref{fig-FL}, a model in a FL methodology has three components from which an attack can be launched: 1) the participants, 2) the server, and 3) the end user. The end user does not participate in training and can only query the shared model. The end user can essentially perform any evasion and exploratory attack that he can launch on general deep models. As far as security is concerned, FL does not open up any additional vulnerability that the end user can exploit. Hence, the evasion and exploratory attacks and defenses on deep models apply directly in FL on the end user side.

The separation of updating server from data introduces more security challenges and attacks. A participant can launch data poisoning or backdoor attack and it is much harder to defend against for two reasons: 1) It is much easier for an attacker to inject and provide data in FL than any other learning strategy by impersonating a participant, and 2) it is much harder for a server to detect poisoned data because it does not have access to the local private data. It is shown that an attacker can can cause the shared model to reach $100\%$ accuracy on the backdoor task in only a single round \cite{bagdasaryan2018backdoor}. Such attacks target integrity of the shared model and it is difficult to defend because the server only observes the parameter updates from participants.

In addition to data poisoning and backdoor attack, a participant can also launch exploratory attacks. For instance, it has been shown that a malicious participant can recover a victim participant's data by training a Generative Adversarial Network (GAN) that generates instances similar to the training data \cite{hitaj2017deep}. This attack is possible because the attacker can update the model weights such that it forces the model to be more over-fitted to the victim's data at each round. The GAN model is trained at each round and gradually learns the distribution of the victim's data. Although exploratory attacks are often known to be effective on over-fitted models, it has been shown that a non-over-fitted model is still prone to generalized membership inference attack \cite{long2018understanding}.
Model extraction attack is trivial in FL because the model is already shared with participants.

The third component of FL is the server. Note that the most paramount goal of FL is to provide privacy of the participants' local data. Hence, an interesting question arises regarding the data privacy: Can the server recover participants' private data only using parameter updates? Due to the separation of training data and updating server, this attack is unique to FL. In this case, the server is assumed honest-but-curious, that is, the server executes the pre-designed training process honestly but may aim to learn or infer private user information. Several papers introduced novel ways to partially recover local private data from participants updates. For instance, if a participant computes the parameter updates using only a single training sample, the server can easily recover the local training data \cite{aono2017privacy}. 
%In \cite{wang2019beyond}, authors introduced a Generative Adversarial Networks (GAN) based model where the discriminator is the shared model and the generator is used to recover participants' private data. %Due to the importance of data privacy in federated learning, several methods have been proposed to defend against such attacks.

The defences against an attacker on end user-side is similar to any defense on general deep learning models and are not unique in FL. What is unique is defence mechanisms to protect privacy of participants' data (against an attacker on server-side or other participants) or integrity of the shared model (against an attacker on participant-side). All defense mechanisms proposed so far present a trade-off among privacy, integrity, computational resources and communication overhead. It is not known whether there is a practical defence mechanism that protects privacy and integrity, and achieve high accuracy or converge in a reasonable time at the same time and much research is needed. In the remained of this section, we discuss defence mechanisms for privacy and integrity separately.

\textbf{Defence mechanisms for privacy:}
There are three main approaches to preserve privacy in FL setting: 1) differential privacy, 2) homomorphic encryption, and 3) Secure Multi-party Computation (SMC). Differential privacy is the most widely used privacy-preserving approach due to its simplicity and theoretical guarantees. The goal of differential privacy is to ensure that the output distribution of the model around a training sample is not too much different from the output distribution of the exact training sample. Over-fitted models that lack generalizability suffer from this overfitting issue and can be a target of the membership inference attack.
The most common approach to guarantee differential privacy is to add small random noise to training data samples. 
%The random noise makes the model more generalizable and more robust to small changes around each training samples. 
It is also possible to add a random noise to the update parameters obtained by participants \cite{geyer2017differentially}. However, there is a trade-off between privacy and accuracy. Increasing the noise value increases the privacy, but it may significantly degrade the accuracy.

The second privacy-preserving approach is homomorphic encryption. Homomorphic encryption allows certain operations on encrypted data in such a way that when decrypted, the results match the results of performing the same operation on the unencrypted data. Homomorphic encryption can be used in two different ways: First, participant can encrypt their data with homomorphic encryption \cite{bost2015machine} and then perform the federated learning as with unencrypted data. In this scenario, even if the malicious server recover a participant's data from the parameters update, it is encrypted and the server cannot decrypt it. However, homomorphic encryption increases computational overhead and also needs polynomial approximation of non-linear functions (such as activation functions), which results in a trade-off between privacy and accuracy. Second, participants can use homomorphic encryption to encrypt their updates. In this case, the server does not have the unencrypted updates to recover the participant's data during the training phase. However, studies that propose using homomorphic encryption on weight updates, such as \cite{aono2017privacy}, use the server as a storage during training. One practical issue with this approach is that the server cannot query the model to observe the progress of the training. Moreover, it is not suitable for the online learning scenario where the model is needed to be used during the training and training is an ongoing process. In such cases, if the unencrypted model is continuously revealed to the server for it to use, it eventually neutralizes the goal of using encrypted weight updates. In such cases, the homomorphic encryption acts similar to SMC, explained in the next paragraph.

%In other words, the server cannot actually use the encrypted weights to respond to queries on the model because it does not have the key to decrypt the weights and use them. Moreover, it only preserve the privacy of training data against the parameter server at best. Hence, at inference time, the model may still prone to exploratory attacks.

The third privacy preserving approach is Secure Multi-party Computation (SMC) \cite{bonawitz2017practical}. Simply put, SMC allows several participants to aggregate their update in a secure way such that the server can only obtain the aggregated updates. Without the exact update of a participant, the server cannot use update to recover local private data. However, SMC cannot still protect the model from information leakage. If the model is inherently prone to data leakage, the attacker does not need individual weight updates. For instance, Hitaj et al. \cite{hitaj2017deep} proposed a Generative Adversarial Network (GAN) model that can recover training samples of participant by actively participating in training phase. Their approach only needs the shared model parameters and works even when SMC is used. The combination of SMC and differential privacy may provide stronger privacy guarantee. However, they are not practical for large-scale scenarios since they incur high computation and communication costs.

\textbf{Defence mechanisms for integrity:}
To defend against malicious participants, most commonly-used mechanisms assume that the majority of participants are honest. Then, they define a metric based on which they can remove malicious updates, keep the most relevant updates, choose the best update, find a robust statistic of the updates, etc. For instance, \cite{yin2018byzantine} uses the coordinated-wise median of all participant updates to update the shared model weights. \textit{Krums} finds the most honest participant at each round and uses its update \cite{blanchard2017machine}. There are several similar approaches that are out of the scope of this article. However, the most important limitation of those approaches is that they implicitly assume an i.i.d. distribution of data among participants which is unrealistic in FL \cite{bagdasaryan2018backdoor}. The impact of non-i.i.d data distribution among participants and how it may affect the detection of malicious updates needs further investigation. Moreover, the performance impact of these approaches where certain number of updates are ignored needs more research. For example, in reality, the number of malicious participants maybe considerably smaller than honest participants. Hence, ignoring a large number of updates, for instance when Krums is used, may degrade the performance and deter fast convergence. Furthermore, certain attacks, such as the backdoor attack, can easily train the local model with similar data distribution as other honest participants and conceal their true purpose. As a result, more research is needed to ensure the integrity of FL.

%Moreover, these approaches are ideal for data poisoning attack where the attacker inject highly irrelevant data. In the backdoor attack, the malicious update can be similar enough to other participants' updates that detecting them with those approaches is unlikely. 

The privacy and integrity aspects of FL seem to be irreconcilable. All current widely-used defences against data poisoning requires unencrypted and unaggregated updates of each participant, making them prone to privacy leakage attacks. On the other hand, all defence mechanisms to preserve privacy manipulate the updates substantially, except for homomorphic encryption on input data, which makes integrity defences useless. To the best of our knowledge, the impact of using homomorphic encryption of input data on weight updates are not well studied. Hence, the weight updates in such cases may be consistent with the assumption of defences for integrity. However, even if the combination of these two achieves good integrity and privacy, the computation overhead and accuracy degradation is not negligible. Therefore, more research is needed to protect the privacy and integrity in FL.

\subsection{Model Compression}
Model compression aims to convert an uncompressed model to a compressed model. The attack target can be either the compressed model or the uncompressed model. If the attack target is the compressed (uncompressed) model and it is available to the attacker, either in a form of black-box or white-box, similar attacks used in deep learning models can be applied and no specific attack is needed. Hence, the more interesting scenario is where an attacker has access to the compressed (uncompressed) model, but she wants to launch the attack on the corresponding uncompressed (compressed) version of the model.

There is only one study on the model compression attack. In \cite{zhao2018compress}, the authors investigate the transferability of adversarial crafted image between compressed and uncompressed models. They study two compression techniques, namely pruning and quantisation, and a few well-known adversarial attacks. Their main observation is that the adversarial inputs can be transferred between compressed and uncompressed model although the attack effectiveness highly depends on the model, the compression method, and the attack algorithm. The adversarial inputs are also transferable between different compressed version of an uncompressed model. Moreover, increasing the compression ratio makes it marginally harder to transfer the adversarial samples.

Since compressed models are often designed for devices directly accessible to customers such as smartphones, it is reasonable to assume that the compressed version is available to an attacker. Hence, the attacker can potentially attack the uncompressed model and all other compressed versions of the model due to the transferability of adversarial samples. Unfortunately, no defense mechanism has been proposed yet. Since the assumptions and the way the attack is launched is somehow similar to TL attacks, attempts to change the compressed version of the model during compression may reduce the transferability of adversarial samples.

The attack and defense mechanisms for the model compression needs further research. For instance, it is not known whether backdoors are also transferable between uncompressed and compressed version. Additionally, the shape of classification manifold may dramatically change when compression is applied. This change may not affect the accuracy of a model on training data, but it may introduce a new possible set of adversarial attacks. It may also be possible for an attacker to meddle the training process of the uncompressed model to make it heavily dependable on all parameters and weights such that it later prevents any compression method to effectively work. %More research is needed to answer these questions.

Exploratory attacks have not been yet studied on model compression. Since the compressed models are usually easy to access from personal devices, it is interesting to study whether the uncompressed model can be extracted from the compressed model. If possible, it would be a threat to businesses that rely on providing compressed models. Similar to TL where the assumption that feature extractor is available opens new attacks, the assumption that compressed model is obtainable may enable more attacks. 

\subsection{Multi-task Learning}
Multi-task learning has been widely used to solve various tasks in image classification, natural language processing, etc. Even when the goal of training a model is to perform single task, we can still train the model for multiple related auxiliary tasks to improve the learning of the target task. However, security vulnerabilities and their potential defense mechanisms have not been studied yet. 

One possible data poisoning attack is to poison the dataset of one task and see if it can be used to target other tasks. Imagine that a victim wants to train a model for facial expression detection. Due to the lack of data, he decides to define an auxilary task of face recognition and uses public datasets. An attacker can poison the public dataset for the auxilary task such that it creates a set of backdoors on the model on which it is trained. How to craft images to create backdoor in this scenario is not a trivial question. 

If an attacker has access to the model, either black-box or white-box, general adversarial attacks on single task models work on multi-task models as well. However, the multi-task model may expose new attacks to the model. For instance, 
lets assume steering direction prediction in self-driving car. A victim may define an auxiliary task of classifying road characteristic and type. Now, since the model is trained on both tasks that are related, the classification output of the road characteristic task probably has a direct relation with the output of steering direction prediction task. By querying the model with different road characteristics, the attacker can find the relationship between these tasks. Although the attacker may not know how to change the input to change the output of the steering direction task, he can change the input to have certain road characteristic that affects the steering prediction task. In other words, an interesting question is that how we can use task \textit{A} to craft an adversarial input targeting task \textit{B}, if task \textit{A} is easier to craft adversarial inputs for. For example, the attacker may have access to a dataset for \textit{A}, but not for task \textit{B}. There are several possible attacks, including exploratory attacks, and potential defenses that are not explored yet.

\subsection{Meta-learning}
In the most general form, we are interested in finding a model or recommender, called \textit{meta-learner}, that outputs a good model or parameters for a new task. For instance, in the transfer learning approach, the meta learner should find the most similar task to the target task and give the suitable pre-trained model. To build such a meta-learner, we often need \textit{meta-data} describing previous tasks, including model architecture and parameters, learning algorithms, and evaluation results. In other words, an entire dataset for task $A$ and an instance of a model that is trained on the dataset is just a single meta-data point for meta-learner. Meta-learner can either recommends a good starting point for training a model for the new task or it can actively learn from models that trained on the new task and iteratively recommend better models. Although meta-learning approaches are vastly different, we discuss them in a general form by considering a general setting using meta-data and meta-learners.

An interesting question is whether meta-learners are vulnerable to any attack. This is not a trivial question, and we have not found any paper regarding meta-learning security. Evasion attack does not seem to be applicable for meta-learning because meta-learners are basically used to train a new model on a new task in an internal training process within a company. In other words, an attacker does not even have an access to the black-box meta-learner to send query to. Even if the attacker hypothetically has an access to the meta-learner, what she can do heavily depends on the meta-learner structure, algorithm, and how it works. Hence, it might be possible to perform an evasion attack on specific meta-learners with a set of non-restrictive assumptions, but it needs more investigation and research.

Data poisoning attack, on the other hand, seems feasible. Here, the attacker should inject a set of meta-data points, not just data points. However, the process of crafting and injecting meta-data point is trickier than data points. In general, data poisoning attack can have two aims: 1) halting or degrading learning process, or 2) creating a backdoor for later use. For the first purpose, the attacker can upload several datasets and models publicly that show good performance, but they do not have suitable model architecture or parameters for the new task meta-learner aims to learn. Although it can slow down the learning process, a good meta-learning strategy should easily defeat this attack by exploring other model architectures and parameters. 

Creating a backdoor attack on the meta-learner alone might not be very interesting to an attacker. The reason is that the attacker probably cannot have access to the mete-learner, as we discussed earlier. Hence, it cannot simply use the backdoor. A more sophisticated and practical attack is to first create a model with a backdoor that achieves a good performance on a task that is not far from the new task meta-learner wants to learn. At the second step, the attacker can force the model to choose his backdoored model, for example as a pre-trained model in the context of transfer learning, by launching data poisoning on the meta-learner. The practicality of such attacks and how to defeat them needs further investigation.

\subsection{Lifelong Machine Learning}
Although lifelong machine learning has been studied for a long time, the security aspect of it has not been comprehensively studied. The reason is that lifelong machine learning is far from being solved and it is still an active area of research. There are a plethora of models and system designs about lifelong machine learning which is out of the scope of this short overview. Here, we only focus on potential security vulnerabilities that a general lifelong learning may have. Each specific lifelong learning model may suffer from additional vulnerabilities, but we only consider general case due to the lack of space. 

Two concepts are highly associated with lifelong learning: 1) the assumption that the previous knowledge is available and it is used to learn new tasks and 2) the sub-goal of preventing catastrophic forgetting. The first assumption enables potential data poisoning, backdoor, and exploratory attacks. The second goal provides a new attack target that aims to break lifelong learning by preventing the system from retaining previous knowledge and tasks, which is an attack on availability.

The study of how backdoor and data poisoning attacks can affect lifelong learning systems is of paramount importance. For example, if a solution manages to tackle catastrophic forgetting, is it possible for an attacker to create a backdoor on one task and use it on all other newer tasks? If possible, it has a disastrous security consequence where all tasks are vulnerable. 

It is also possible to jeopardize the second sub-goal, i.e. solving catastrophic forgetting. One interesting approach is to study whether adding a few carefully crafted training samples, with correct labels with respect to the new task, can change the structure of the model such that it performs poorly on older tasks. An attacker can potentially formulate an optimization problem where the goal is to change the manifold of the old task dramatically while the label is correct with respect to the new task. The attack and defense mechanisms specific to the lifelong learning need more research.

\section{CONCLUSION}
The adaptation of different machine learning methodologies has significantly growth recently and their security vulnerabilities should be studied comprehensively.
In this article, we briefly introduce several widely-used machine learning methodologies and discuss their potential security vulnerabilities and challenges. We note that the security aspects of certain machine learning methodologies are open research questions that call for further investigation. We hope that this article enlighten the machine learning community about the potential security vulnerabilities of these methodologies and encourage more research to prevent any backlash as a result of security breakdown.

\begin{IEEEbiography}{Shahbaz Rezaei}{\,} 
received his M.S. degree in information technology from the Sharif University of Technology, Tehran, Iran, in 2013. His research interests include machine learning security, machine learning application, deep reinforcement learning, and computer networks. He is currently a Ph.D. student at UC Davis. Contact him at srezaei@ucdavis.edu.
\end{IEEEbiography}

\begin{IEEEbiography}{Xin Liu}{\,}
received her Ph.D. degree from Purdue University in 2002. She is currently a professor in the Computer Science Department, University of California, Davis. Her current research focuses on data-driven approach in networking  (i.e., using and developing machine learning and optimization techniques for network control and management). She is an IEEE Fellow. Contact her at xinliu@ucdavis.edu.

\end{IEEEbiography}

\end{document}